\documentclass[prl, reprint, superscriptaddress,groupedaddress, unsortedaddress,nofootinbib, amsmath,amssymb,aps,showkeys,floatfix,
]{revtex4-2}

\usepackage{graphicx}
\usepackage{dcolumn}
\usepackage{bm}
\usepackage{physics}
\usepackage{comment}
\usepackage{tabularx}
\usepackage{color}
\usepackage{hyperref} 
\hypersetup{
colorlinks=true,
linkcolor=blue,
linktoc=page,
citecolor=blue,
urlcolor=blue
}

\everymath{\displaystyle} 

\begin{document}

\preprint{APS/123-QED}

\title{Superdiffusion of vortices in two-component quantum fluids of light}

\author{M. Rold\~{a}o}%
 \email{miguel.roldao@tecnico.ulisboa.pt}
\author{J. L. Figueiredo}%
\author{P. Monteiro}%
 \author{J. T. Mendon\c{c}a}
\author{H. Ter\c{c}as}
 \email{hugo.tercas@tecnico.ulisboa.pt}
\affiliation{%
 GoLP | Instituto de Plasmas e Fus\~{a}o Nuclear, 
 Instituto Superior T\'ecnico, Universidade de Lisboa, Portugal}%

\begin{abstract}
\par The quantum diffusion of a vortex in a two-component quantum fluid of light is investigated.  In these systems, the Kerr nonlinearity promotes interactions between the photons, displaying features that are analogue of a Bose-Einstein condensates. Quantum fluids of light have the advantage of simulating matter-wave phenomena at room temperatures. While the analogy is true at the mean field level, the full quantum dynamics of an impurity in quantum fluids of light of, and therefore the ability of featuring genuine quantum noise, has never been considered. We numerically solve the problem by simulating a vortex-like impurity in the presence of noise with the Bogoliubov spectral density, and show that the vortex undergoes superdiffusion. We support our results with a theory that has been previously developed for the brownian motion of point-like particles.
\end{abstract}
\keywords{Quantum Fluids of Light, Quantum Polaron, Brownian Motion, Superdiffusion, Quantum Vortex.}
\maketitle

{\it Introduction.---} The first observations of diffusive processes in nature are due to Brown in 1827, who reported on the erratic motion of wildflower pollen suspended in water \cite{brown1828xxvii, erdos_2010}. Theoretical understanding of the brownian motion remained poor until 1905, with Einstein's and Smoluchowski's kinetic explanations, proposing that the heavy pollen particules are propelled by the sucessive kicks by the light water molecules \cite{einstein1905molekularkinetischen, smoluchowski1906kinetic}. The success of the kinetic theory of the brownian motion led to Perrin's estimation of Avogadro's number in 1908 \cite{bian_2016}. An intriguing variation of the phenomenon appears in the quantum regime \cite{Meyer_1996, lampo_2019}, with profound consequences in condensed matter systems, such as Anderson localization, which has been observed for both matter and light waves. Quantum brownian motion may originate from random external potentials such as disorder or via the coupling to a random reservoir. In the latter case, the Caldeira-Legget model provides a semi-empirical treatment of the coupling between a quantum particle and the environment, overcoming the difficulties of obtaining a Langevin equation from the standard procedures of quantization \cite{caldeira_1983a, caldeira_1983b, caldeira_1993}. Depending on the specific form of the particle-bath coupling, the system may display normal or anomalous diffusion \cite{bouchad_1990, hegde_2022}. Over the recent years, a great deal of attention has been given to the investigation of quantum diffusion in cold atoms, with the intention of implementing the polaron problem by immersing quantum impurities in a Bose-Einstein condensate (BEC) \cite{zipkes_2010,schmid_2010, catani_2012, spethmann_2012, scelle_2013, rentrop_2016, schmidt_2018}. Recently, the quantum diffusion of a Bose polaron in BECs has been suggested as a route towards high-sensitivity measurements of temperature in the sub-nanokelvin regime \cite{miskeen_2021, miskeen_2022}. \par
Another interesting physical system in which quantum brownian motion could possibly occur, albeit lacking investigation, are quantum fluids of light (QFL), often referred as the optical analogues of Bose-Einstein condensates \cite{carusotto_2013, carusotto_2014}. Quantum fluids of light are platforms where photon-photon interactions are possible thanks to the Kerr nonlinearity, encompassed in the nonlinear part of the susceptibility, $\chi^{(3)}$. The fluidic characteristic of a light beam has been first discussed by Brambilla \cite{Brambilla} and Staliunas \cite{LaserStaliunas}, but have received a great deal of attention in the more recent years \cite{joyce_1999, larre_2015, vocke_2015, michel_2018, ferreira_2018}. At the mean-field level, paraxial QFLs are governed by an equation of the Ginzburg-Landau type, which contains a biunivocal correspondence with the Gross-Pitaevskii equation \cite{carusotto_2014}. Such correspondence has allowed the emulation many physical effects taking place in BECs, such as solitons and vortex formation \cite{swartzlander_1992}, quantum turbulence \cite{rodrigues_2020, silva_2017, Silva_2021, ferreira_2022, rasooli_2023} and spectral condensation \cite{sun_2012, selse_2018}, Rayleigh-Taylor instabilities \cite{jia_2012}, and have even been pointed as analogue (also known as hydrodynamic, or Unruh \cite{unruh_1981}) black holes linked to the existence of a local sound speed \cite{marino_2008, solnyshkov_2011, nguyen_2015, carusotto_book, jacquet_2023}. Moreover, spin-orbit coupling in QFLs are possible in nonparaxial configurations, making possible the quantum simulation with multi-component BECs \cite{martone_2021}.
\begin{figure}[!t]
    \centering
    \includegraphics[width=\columnwidth]{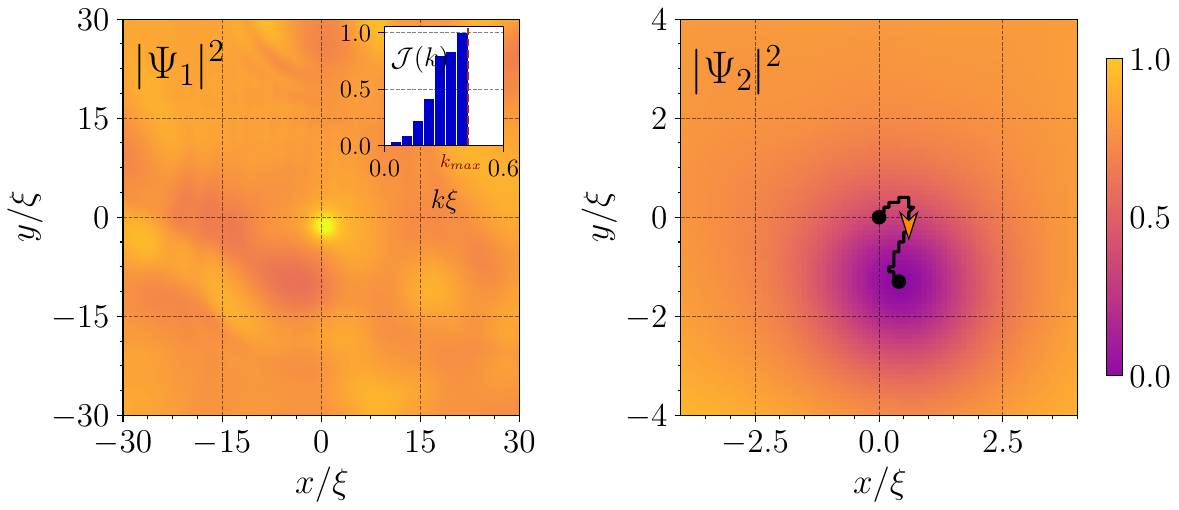}
    \caption{(color online): Brownian motion of a vortex in a quantum fluid of light with two orthogonal polarizations. Left panel : The bath excitations are introduced in the first polarization of the electric field and are randomly initiated according to the Bogoliubov spectral density, $\mathcal{J}(k)$ in the inset. The cutoff wavevector has been chosen to verify the condition $\lambda_{\text{min}}=2\pi/k_{\rm max}=16\xi$. Right panel: The vortex in the second component plays the role of a quantum impurity. It is initially located at the center $(x,y)=(0,0) $ and is propelled by the action of the noise. In both panels, we have set $t=25\tau$.}
    \label{fig:ITP_BEC}
  \end{figure}
Another crucial benchmark of the viability of QFLs as quantum simulators of BECs has been the recent observation of the Bogoliubov spectrum, an aspect that is intimately related to the bosonic nature of the quantum excitations \cite{fontaine_2018, piekarski_2021}. Therefore, testing the ability of the latter to produce superdiffusion would constitute a critical test to the full quantum mechanical nature of QFLs, therefore opening the door towards the simulation of open quantum systems with fully optical setups.
\par
In this Letter, we theoretically investigate the brownian motion of a quantum vortex in a two-component quantum fluid of light propagating in a rubidium vapor, which is an appealing platform for applications quantum technologies \cite{quantum.memories.with.cesium, Katz_Firstenberg_2018}. In our scheme, the two components correspond to the different polarizations of the electric field \cite{Silva_2021}. We show that a quantum vortex undergoes {\it superdiffusion}, which is compatible with the superohmic nature of the Bogoliubov spectrum. 
\par {\it Two-component quantum fluids of light.---} Quantum fluids of light (QFL) refer to the collective behavior of photons in nonlinear media. Considering a centrosymmetric media, such as alkali metals, the polarization field $\bm{P}$ can be expanded to the third-order in the electric field $\bm{E}$  as
\begin{equation}
    \bm{P} = \epsilon_0 \left( \chi^{(1)}\bm{E} + \chi^{(3)}|\bm E|^2\bm{E} \right) + \mathcal{O}(E^5),
\end{equation} 
where $\epsilon_0$ is the dielectric permittivity and $\chi^{(1)}=\sqrt{n_1^2-1}$, with $n_1=c/(\partial_k\omega_L)$ is the (linear) refractive index. The term $\chi^{(3)}$ is commonly designated as the nonlinear optical Kerr effect \cite{mansuripur_2009} and leads to a dependence of the index of refraction on the applied electric field, i.e. the intensity of the photon beam. For rubidium, the Kerr susceptibility $\chi^{(3)}={\rm Re}\chi^{(3)}(\delta, T)+i \rm{Im}\chi^{(3)}(\delta, T)$ is a function of the detuning $\delta=\omega_L-\omega_0$, where $\omega_L$ is the laser frequency and $\omega_0$ is the D2-transition frequency and $T$ is the temperature \cite{aladjidi_2022}. In what follows, we decompose the electric field into two orthogonal polarizations $\bm{\varepsilon}_\alpha$, with $\alpha=\{1,2\}$,  as $\bm{E}(\bm r, t) =\sum_\alpha E_\alpha(\bm{r}_\perp,t)\bm \varepsilon_\alpha e^{ik_Lz-i\omega_L t}$, with $\bm{r}_\perp=(x,y)$ denoting the perpendicular direction and $z$ the propagation direction. Under the paraxial approximation, each polarization is governed by the following equation \cite{Kivshar, SILVA2021100025}
\begin{equation}
      i\dfrac{\partial E_\alpha}{\partial z} = -\frac{1}{2k_L}\nabla_{\perp}^2 E_\alpha - \frac{3}{2} n_2 k_L|E_\alpha|^2 E_\alpha - n_2 k_L|E_\beta|^2 E_\alpha,
      \label{eq_paraxial}
\end{equation}
where the nonlinear refractive index is given by $n_2 \equiv \chi^{(3)}/n_0$, with $n_0$ being the rubidium vapor density (which, in turn, is controlled through the temperature $T$). Recent quantitative measurements show that its absolute value $|n_2|$ increase with the vapour density and decrease with both $\vert \delta\vert$ and beam intensity $\vert \bm{E}\vert^2$ \cite{Wang:20}. As $n_2$ is also frequency dependent, the frequency detuning $\delta$ may be such that $n_2<0$, which corresponds to a self-defocusing medium. This is the conditions we consider in the remainder of this manuscript. \par

For a more quantitative analysis, it is convenient to map the optical problem into a Bose-Einstein condensate. In the zero temperature limit, BECs can be described by a macroscopic order parameter $\Psi$ that obeys the Gross-Pitaevskii equation. This phenomenon of condensation involves a dominant occupation of the lowest quantum state, which motivates writing the $\hat{\Phi}$ operator as a sum of two terms, $\hat{\Phi}(\bm{r},t) = \Psi(\bm{r},t) + \delta\hat{\varphi}(\bm{r},t)$. The first term $\Psi(\bm{r},t)= \bra{0}\hat{\Phi}(\bm{r},t)\ket{0}$ is the mean field order parameter and refers to the condensed fraction of the system, while $\delta\hat{\varphi}(\bm{r},t)$ corresponds to the fluctuations (or excitations), the fraction that is not condensed in the GS. Since we are considering a weakly interacting gas ($a_s \ll \lambda_{dB}$, the de Broglie wavelength), excitations are expected to be small. Neglecting the excitations, we have the Gross-Pitaevskii equation for the condensate
\begin{equation}
\label{eq:GPE_final_BEC}
    i\hbar \dfrac{\partial\Psi}{\partial t} = \left[-\dfrac{\hbar^2\nabla^2}{2m} + V_{\text{trap}}(\bm{r}) + g |\Psi|^2\right] \Psi.
\end{equation}
This equation is valid for Bose-Einstein condensates if the temperature of the system is much lower than the critical temperature of condensation, $T_c$. For these systems, the Bogoliubov dispersion relation is written as 
\begin{equation}
\label{eq:disp_relation_BEC}
    \omega^2 = c^2k^2 + \frac{\hbar^2k^4}{4m^2},
\end{equation}
where $c_s = \sqrt{gn_0/m} = \xi/\tau$ is the speed of sound \cite{bogoliubov1947theory}. Equations \eqref{eq_paraxial} and \eqref{eq:GPE_final_BEC} have strong similarities between their terms, this being what justifies the parallelism established between nonlinear optics and the BEC physics. \par
{\it Diffusion and non-Markovianity.---} We intend to simulate the quantum dynamics of an impurity in a Bose-Einstein condensate. This system is ruled by a system of coupled Gross-Pitaevskii equations. For quantum fluids of light, interactions between bosons have a fixed relation of $\sigma \equiv g_{21}/g_{11} = 2/3$ \cite{Kivshar,SILVA2021100025}; space and time are scaled such that $\tau = t/t' = \hbar/\mu$, $\xi_i = r/r' = \hbar/\sqrt{g_{ii}m_in_0}$ and $g_{ii} = \mu_i/n_0$ with $i=\{1,2\}$. We are interested in the case $m_1$=$m_2$ and $g_{11}=g_{22}$, such that Eqs. \eqref{eq_paraxial} read
\begin{equation}
        \begin{cases}
            i\dfrac{\partial\Psi_1}{\partial t'} = \left[- \dfrac{1}{2} \nabla_\perp^2 + |\Psi_1|^2 + \sigma |\Psi_2|^2 \right]\Psi_1\\
            \\
            i\dfrac{\partial\Psi_2}{\partial t'} = \left[- \dfrac{1}{2} \nabla_\perp^2 + \sigma|\Psi_1|^2 + |\Psi_2|^2\right] \Psi_2.
        \end{cases}
        \label{eq_caceta}
    \end{equation}
To solve the system in Eq. \eqref{eq_caceta}, we follow a finite-difference, Cranck-Nicholson scheme. A Runge-Kutta method of $4^{\text{th}}$ order in the time domain is used. The simulation parameters were chosen to guarantee convergence \cite{leveque1998finite} ($dx^2/dt \geq 2$), $dx=0.1\xi$, $dt=0.005\tau$. The wavefunction $\Psi_1$ is the background that holds the bosonic excitations, while $\Psi_2$ is the impurity, a vortex-like defect in density, which we initialize with the help of a imaginary-time evolution method. Dirichlet boundary conditions were used with a squared box of side $120\xi.$ The thermal bath corresponding to the Bogoliubov excitations has the prescribed spectral density. The cutoff frequency $\Lambda_c$, or the $\lambda_{\text{min}}$, selected the dominant frequency in the noise spectrum. A snapshot of the simulation can be found in Fig. \ref{fig:ITP_BEC}. 
\begin{figure}[!t]
    \includegraphics[width=0.4\textwidth]{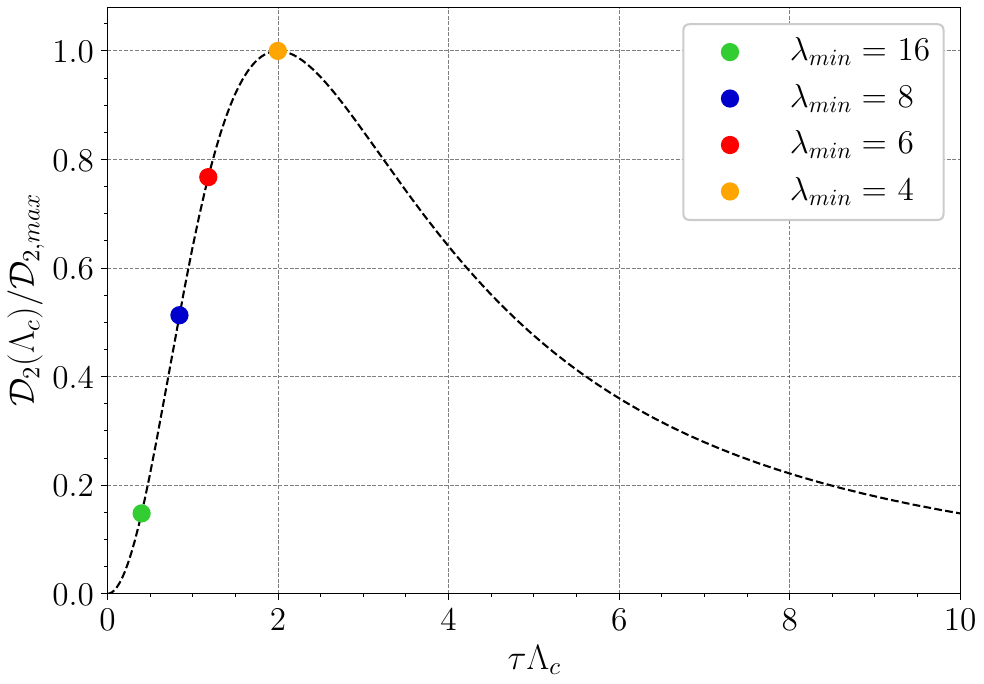}
    \caption{(color online) Diffusion coefficient $\mathcal{D}$ as a function of the cutoff frequency $\Lambda_c$. For a point-like impurity, we find a maximum for $\tau \Lambda_c=2$.}
    \label{fig:MSD_predict}
\end{figure}
\par
As described in \cite{miskeen_2021}, the dynamics of an impurity in a BEC implies the interactions between this particle and the thermal heat bath where it is immersed. The total Hamiltonian is 
\begin{equation}
  H = H_I + H_B + H_{BB} + H_{IB}.
\end{equation} 
The spectral density of the heat bath is calculated through the tensor $\underline{\underline{g_{\boldsymbol{k}}}}=\underline{g_{\boldsymbol{k}}}\cdot \underline{g_{\boldsymbol{k}}}^T$ as
\begin{equation}
\label{eq:Spec_dens_delta}
    \underline{\underline{J}}(\omega)=\sum\limits_{\boldsymbol{k}\neq 0}\hbar \underline{\underline{g_{\boldsymbol{k}}}}\delta(\omega-\omega_{\boldsymbol{k}}).
\end{equation}
For the generic d-dimensional case the spectral density is a diagonal tensor with entries $\mathcal{J}_d$ (due to spatial symmetry of the noise). For a point-like particle in two dimensions (QFL analogy),
\begin{equation}
\label{eq:Spec_density}
    \mathcal{J}_2 (\omega) = \frac{\sigma^2 \omega_0^4 m_B^2}{g_B n_0^2 \sqrt{\frac{\omega^2}{\omega_0^2}+1}}\left(\sqrt{\frac{\omega^2}{\omega_0^2}+1}-1 \right)^2.
\end{equation}
Here $m_B$ is the bosonic mass, $n_0$ is the untrapped density, and $\omega_0 = \tau^{-1} = \mu/\hbar$. Other important aspect of the impurity dynamics is due to the Markovianity of its motion. Formally, Markovianity is tackled through the Langevin equation for the motion of the impurity
where the damping kernel $\Gamma(t)$ is obtained from the Spectral density $\underline{\underline{J}}$ as
\begin{equation}
\label{eq:Damping_kernel}
    \underline{\underline{\Gamma}}(t) = \frac{1}{m_I}\int_0^\infty d \omega \frac{1}{\omega}\;\underline{\underline{J}}(\omega) \cos(\omega t).
\end{equation}
When the damping kernel is not zero, the dynamics of the system has a memory term. Only when the damping kernel is zero the dynamics is Markovian: in terms of the Spectral density, when $J(\omega) =\eta \omega$.
\par The mean square displacement of the impurity (MSD), for the long time limit in the two dimensional case, is
\begin{equation}
\label{eq:MSD_expression}
    \langle [x(t)-x(0)]^2 \rangle = \left[ \langle \dot x(0)^2\rangle + \frac{\hbar \tau^2\Lambda_c^3}{6m_I}\right]\left(\frac{t}{\alpha_2}\right)^2,
\end{equation}
where $\alpha_2=1+\frac{1}{4}(\Lambda_c \tau)^2$.
Considering the characteristic healing length $\xi=\hbar/\sqrt{\mu m}$ and $\tau=\hbar/\mu$, being the chemical potential $\mu=g n_0$, we get
\begin{equation}
   \frac{\text{MSD}(t)}{\xi^2} = \frac{m_B}{m_I} \frac{\tau \Lambda_c^3}{6}\left(\frac{t}{\alpha_2} \right)^2,
\end{equation}
choosing $\langle \dot x(0)^2\rangle=0$. The ratio $m_B/m_I$ influences the amplitude of the movement, such as the cutoff frequency $\Lambda_c$. A superdiffusion regime have therefore been predicted and the system should contain non-Markovian effects for a Bogoliubov heat bath acting on an impurity \cite{miskeen_2021}. The diffusion coefficient is defined as $\mathcal{D}_2 = \hbar \tau^2\Lambda_c^3/(6 m_I\alpha_2^2)$.
Comparing with the numerical results we assessed if the behaviour of the vortex was similar to the one of the localized particle or if their different wavefunction distributions are relevant. In Fig. \ref{fig:MSD_predict}, the non-monotony of the diffusion coefficient, with respect to $\Lambda_c$, is shown. We should remind ourselves that these results were obtained for a localized impurity. 
\par
\begin{figure}[!t]
    \includegraphics[width=\columnwidth]{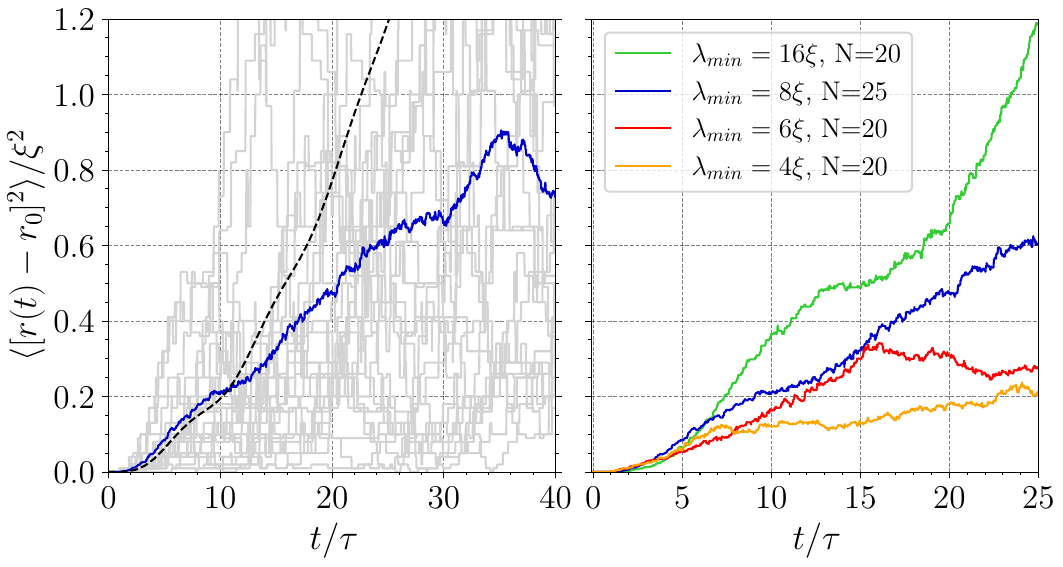}
    \caption{(color online) Superdiffusion of a quantum vortice driven by the quantum noise. Left panel: the comparison between the predicted behaviour of mean-square displacement (MSD) (black dashed curve) and the results for $\lambda_{\text{min}}=8\xi$ - grey lines are the individual runs, the blue line is the averaged MSD. Right panel: Vortex mean-square displacements for different values of the cutoff.}
    \label{fig:MSD}
\end{figure}

 We may integrate the expression for the MSD numerically to predict the behavior of an impurity in the quantum bath. In Fig. \ref{fig:MSD_predict}, we see the behaviour of the nonlinear diffusion coefficient $\mathcal{D}$, such as the values of $\lambda_{\text{min}}$ used in the simulation. To observe the effect of $\Lambda_c$, the simulation was run with 4 different values (table \ref{tab:runs_details}). As it is shown in figure \ref{fig:MSD}, the initial growth of the MSD is similar to the prediction, but started diverging after $t=10\tau$. The analytical prediction for the MSD is in disagreement with our results by a scaling factor of 200. This can be related to the mass of the vortex being distinct to the bosonic mass, therefore larger than the previously considered point-like impurity.
\begin{table}[!t]
    \renewcommand{\arraystretch}{1.2} 
    \centering
    \begin{tabular}{m{2cm} m{2cm} m{2cm} m{2cm}}
    \hline
    \hline
    \centering $\lambda_{\text{min}}[\xi]$ & $\Lambda_c^{-1}[\tau]$ & $N_{\text{runs}}$ & $t_{\text{max}}[\tau]$ \\ \hline
   \centering 16 & 2.50 & 20 & 25 \\
   \centering 8 & 1.18 & 25 & 40 \\
   \centering 6 & 0.85 & 20 & 25 \\
   \centering 4 & 0.50 & 20 & 25\\ 
    \hline
    \hline
    \end{tabular}
    \caption{Parameters of the simulation in its characteristic units $\xi=\hbar/\sqrt{\mu m}$ and $\tau=\hbar/\mu$.}
    \label{tab:runs_details}
\end{table}
\par An oscillation with a period of $\sim 10 \tau$ is present in the results in accord with the theoretical model \cite{miskeen_2021}. It predicts that this oscillation becomes negligible for large time scales as a quadratic dependency in time emerges (see Eq. \ref{eq:MSD_expression}). The simulation results do not confirm this quadratic dependency, as time increased. Comparing the different $\lambda_{\text{min}}$ runs, it is clear that the MSD increases in general for larger values of this parameter, being $\lambda_{\text{min}}=16$ the largest value that was tested. Aiming to simulate an untrapped impurity with Dirichlet boundary conditions, the $\lambda_{\text{min}}$ was kept below a size comparable with $60\xi = (x_{\text{max}}-x_{\text{min}})/2$, to prevent boundary effects.
\par The relation between $\mathcal{D}$ and $\Lambda_c$ is not in agreement with the theoretical prediction (see figures \ref{fig:MSD_predict} and \ref{fig:MSD}). While the prediction for this range of $\Lambda_c$, the diffusion coefficient should be increasing with $\Lambda_c$. The simulation results indicate otherwise, as we would be on the decreasing regime of the curve in \ref{fig:MSD_predict} ($\tau \Lambda_c >2$). 
\par
\begin{figure}[!t]
    \includegraphics[width=\columnwidth]{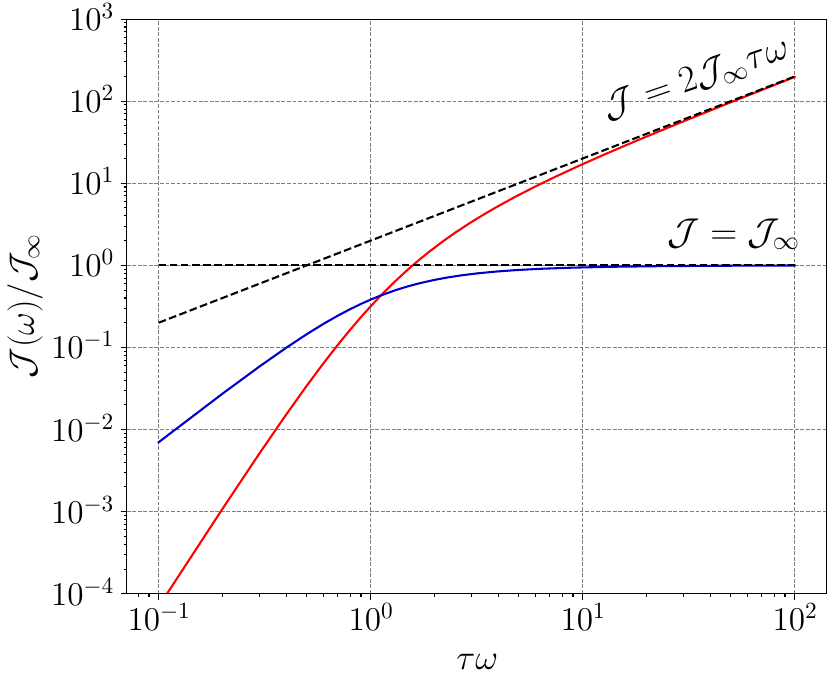}
    \caption{(color online) Particle (red) and vortex (blue) spectral densities diagonal element. $\mathcal{J}_{\infty}=n_0g_{IB}^2/2h\xi$.}
    \label{fig:Spec_density_new}
\end{figure}

 These disagreements suggest that the behaviour of a vortex in the heat bath is quite distinct from the prediction for a point-like particle. Therefore the spectral density was recalculated following the same method of Ref. \cite{miskeen_2021}, with a vortex-like density profile
\begin{equation}
    \rho_I(\boldsymbol{q}) = \int_{-\infty}^\infty d\boldsymbol{r}' e^{-i\boldsymbol{q}\cdot \boldsymbol{r}'}\delta(\boldsymbol{r}'-\boldsymbol{r}) \frac{(\boldsymbol{r}-\boldsymbol{r'})^2}{\sqrt{(\boldsymbol{r}-\boldsymbol{r'})^2+ \xi^2}}.
\end{equation}
This change in the radial density of the vortex with an aperture of the order of the healing length $\xi$, gives an expression for the spectral density such as
\begin{equation}
\label{eq:spec_vortex}
    \mathcal{J}(\omega) = \frac{n_0 g_{IB}^2}{2 h\xi} \frac{\omega/\Lambda_c \left(\sqrt{1+\frac{\omega^2}{\Lambda_c^2}}-1\right)^{1/2}}{\left( 1+\frac{\omega^2}{\Lambda_c^2}\right)^{3/4}},
\end{equation}
different from Eq. \ref{eq:Spec_density}. This small change changes the influence of the excitations on the impurity dynamics. In Fig. \ref{fig:Spec_density_new}, the different behaviour of the coupling coefficient $|g_{\boldsymbol{k}}|$ and spectral density diagonal elements $\mathcal{J}$ are shown. While for a particle, as far as the frequency $\omega$ increase, the coupling increases, and the higher frequency dominates, for a localized vortex, the interactions saturate to a fixed value $\mathcal{J}_{\infty}$. \par
As we increase the excitations frequency $\omega$, the relevance of each frequency to the vortex stabilizes and all the frequencies tend to have the same importance. This means none will resonate to displace the vortex, as we can see from Eq. \ref{eq:Spec_dens_delta}. A particle will always have a higher $\omega$ to the wandering motion so the diffusive coefficient should have a behaviour with $\Lambda_c$ such as shown in figure \ref{fig:MSD_predict}. In these systems, we observe superdiffusive motion, as the Bogoliubov spectrum of the quantum noise selects specific frequencies to be dominant in the spectrum. The Bogoliubov spectrum is not homogeneous. Ultimately, looking at equation \ref{eq:Damping_kernel}, superdiffusion emerges from the fact that this resonant behaviour between the quantum noise and the impurity, is not exactly random, in the Brownian sense, but is responsible for this faster diffusive effect, which has the characteristic fingerprint of a time dependence such as MSD$\sim t^\alpha$, with $\alpha>1$ the superdiffusion index. Of course, this superdiffusion is highly dependent on the noise spectrum. If we introduce white noise to the system, we expect that the damping kernel does not select any specific frequency and the memory term in Langevin equation disappears and we recover the Brownian motion. This is also what happens in the high temperature regime $\hbar\omega/k_B T\xrightarrow{}0$, as we lose the memory effects. These conclusions are deeply connected to the markovianity of the system. The noise works as a memory external bath through the noise kernel $\underline{\underline{\Gamma}}(t)$, so the markovianity of the system is highly excitation-dependent, i.e. dependent on the spectrum of the bath.
\par The accuracy of the mean-squared displacement, being a statistical quantity, is increased with the number of runs. Its error will increase in time for diffusive processes and will require more runs to predict more accurately its behaviour for larger time scales.
\vspace{3mm}
\par
{\it Conclusions and future work.---} Quantum fluids of light were introduced as an analogy between nonlinear optics and Bose-Einstein condensation, a formal analogy that may be of extreme importance and relevance in both areas, as well as for real quantum computing applications, such as the development of quantum memories. Regarding the study of diffusive effects on Bose-Einstein condensates, a numerical simulation involving the interaction between a vortex solution, as an impurity and a thermal Bogoliubov bath was studied. All the parameters were discussed, as well as the important and relevant details from the numerical point of view, such as the conditions to achieve numerical stability, but also physically, for example, the imaginary time evolution procedure. The results were analyzed and motivated a deeper study of the assumptions that were made, leading to a detailed discussion of the impurity density profile and its role in the interactions. The regimes of superdiffusion or standard Brownian diffusion were evaluated and their relation with the loss of markovianity was presented as a direct consequence of the interactions of the impurity with the quantum noise bath, in particular, a direct consequence of the noise spectrum that is used. The spectrum of the noise was discussed to have immense importance to the behaviour of the system and its super or subdiffusive regime. We concluded that in the situation where a white noise spectrum is included, or we reach the high-temperature regime, superdiffusion is lost, as well as the memory effects intrinsic to non-markovianity. Aiming to reach more diverse results in this matter, the next steps were discussed as well as other interesting configurations that might be studied with the same algorithm. \par
{\it Acknowledgements.---} The authors acknowledge the financial support from Fundac\c{c}\~{a}o para a Ci\^{e}ncia e a Tecnologia (FCT-Portugal) through Contract No. CEECIND/00401/2018, Project No. PTDC/FIS-OUT/3882/2020, and Grant No UI/BD/151557/2021.

\bibliography{apssamp}
\bibliographystyle{apsrev4-2}

\end{document}